# A Probability Density Function for Google's stocks

V.Dorobantu

Physics Department,
"Politehnica" University of Timisoara, Romania

*Abstract.* It is an approach to introduce the Fokker – Planck equation as an interesting natural ingredient in studying the evolution of the market stock prices.

**Introduction.**

From physical point of view, in many cases, we are interested to study the behavior of a certain system – S. That system may be coupled to another system, let it be a reservoir – R, or may be not. If we have to do with coupling, then our system "feels" the influence of the reservoir so, we have to consider the behavior of the entire system S+R. Finally, being interested in properties of the system S only, it is necessary to find a way to eliminate the variables of R. A thorough analysis has been done by Haken [1] showing that an equation of motion of the system S coupled linearly to the reservoir R is of the Langevin type:

$$\partial_t x(t) = D(x(t), t) + f(t) \tag{1}$$

where x is a variable (or a set of) of our system S, D some time – dependent "force" and f(t) a fluctuating "force" coming from reservoir R. One can show [1] that a random "force" f(t) can be taken as:

$$f(t) = s \sum_j (-1)^{n_j} \delta(t - t_j) \tag{2}$$

with s being the size of the impulses, $n_j$ being zero or 1 depending on the fact that the random impulses are in forward or backward direction and $\delta(t - t_j)$, the Dirac's function.

What we need is the correlation function $\langle f(t) f(t') \rangle$ which can be taken [1] as:

$$\langle f(t) f(t') \rangle = C \delta(t - t') \tag{3}$$

with C a constant and $\langle \ \rangle$ the statistical average. A straightforward continuation is the Fokker – Planck equation:

$$\partial_t \Psi(x, t | x_0, t_0) = -\partial_x (D_1(x) \Psi(x, t | x_0, t_0)) + \frac{1}{2} \partial_{x,x} (D_2(x) \Psi(x, t | x_0, t_0)) \tag{4}$$

where $\Psi(x,t|x_0,t_0)$ is the probability density function (pdf), that is the probability to have the variable x at time t if it had the value $x_0$ at a preceding time $t_0$. In fact, $\Psi(x,t|x_0,t_0)$ can be taken as $\Psi(x,t)$, namely the probability to have x in the range x and x + dx at a time t. $D_1(x)$ is the drift coefficient defined as:

$$D_1(x) = \lim_{t \to 0} \frac{\langle x(t) - x_0 \rangle}{t} \tag{5}$$

and $D_2(x)$ the diffusion coefficient:

$$D_2(x) = \lim_{t \to 0} \frac{\langle x(t) - x_0 \rangle^2}{t} \tag{6}$$

**The stocks market and the Fokker – Planck equation**

It seems to be very natural to consider that if at a certain time $t_0$ the price of a stock being $x_0$, to ask what it will be at the subsequent time t, and the answer to this question to be done by the a probability density function satisfying a Fokker – Planck equation.
If we look at Google's stock prices [2], the stock price, x(t), can be considered as a continuous variable, and as a random one, also. Nevertheless, the "reservoir" may have an influence upon prices: a favorable (or not) article in a very known newspaper, an appreciation of a consulting company, a political interest at a certain moment of time, etc. So, the fluctuating "force" done by (2) can have $n_j = 0$, or $n_j = 1$, and as a consequence, the correlation function is of the form done by (3).

**First version, linear dependence**

Following Haken [1], we are in a *thermodynamics* approach of a Brownian motion where the "coherent force" $D_1(x)$ can have a certain expression and $D_2(x) = C$. Let us take C = 1 as a hypothesis. The time evolution of Google's stocks price has a, rather, complicated dependence on time, and it is not specific for Google, it's generally valid. In the limits of experimental data, as physicists say, we can simplify that dependence assuming a linear one. The stock price, x(t), can be taken as:

$$x(t) = a + bt \tag{7}$$

Allowing x(t) being of the form (7), $D_1(x) = b$, and reminding $D_2(x) = 1$, the Fokker – Planck equation takes the form:

$$\partial_t \Psi(x,t) = -b \partial_x \Psi(x,t) + \frac{1}{2} \partial_{x,x} \Psi(x,t) \tag{8}$$

The solution of the above equation (8) is:

$$\Psi(x,t) = \frac{1}{\sqrt{2\pi t}} e^{-\frac{(x-a-bt)^2}{2t}} \qquad (9)$$

Using data [2] regarding close and high stock's prices, an average between them, x(t) can be taken as:

$$x(t) = 129 + 0.63\, t \qquad (10)$$

Fig. 1 shows such a dependence and the confidence interval, as well. t is the day's number, namely 1 for the first day (Aug.19, 2004), 2 for the second working day and so on. The data cover the interval between t = 1 (Aug.19, 2004), t = 573 (Nov.24. 2006).

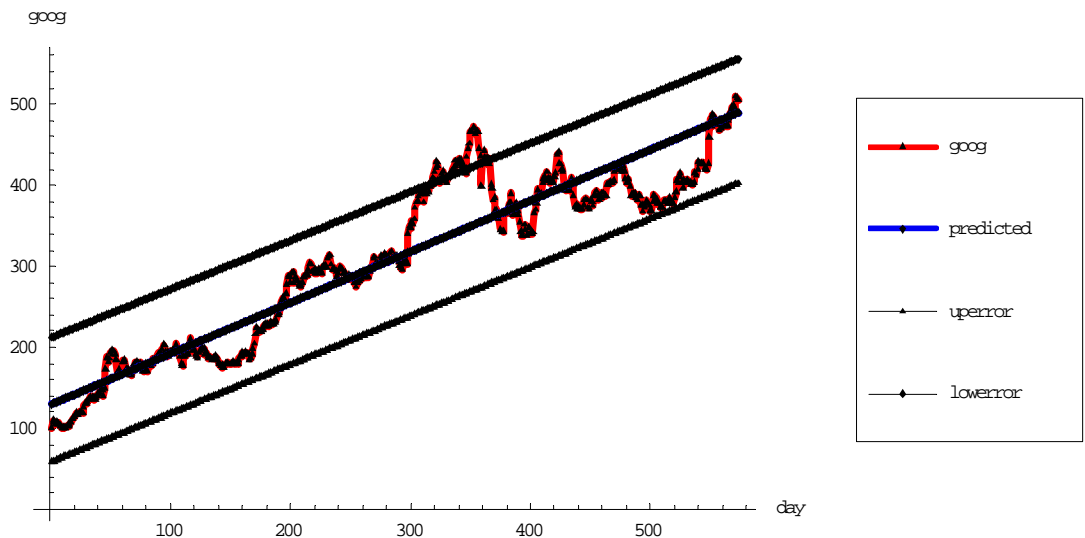

Fig.1

**Second version, quadratic dependence**

What will happen if the stock's price will reach a maximum value and then starts to decrease? Obviously, the time evolution shall not have the form done by (7). Processing the data concerning the Google's prices at close session, we can get an expression like this one:

$$x(t) = a\,t^2 + b\,t + c \qquad (11)$$

Such a form will change the drift coefficient $D_1$, and $D_2$ as well

$$D_1(x,t) = 2at + b \qquad D_2(x,t) = 1 + (C_1 + x\,C_2)\, e^{\frac{x(x+2at^2-2bt-2c)}{2t}} \qquad (12)$$

With (12), the solution of the Fokker-Planck (4) is:

$$\Psi(x,t) = \frac{1}{\sqrt{2\pi t}} e^{-\frac{(x-c-bt-at^2)^2}{2t}} \quad (13)$$

Using data [2] regarding close stock's prices, x(t) can be taken as:

$$x(t) = -0.00013\, t^2 + 0.65\, t + 135 \quad (14)$$

Fig. 2 shows the quadratic dependence and the confidence interval, as well. t is the day's number, namely 1 for the first day (Aug.19, 2004), 2 for the second working day and so on. The data cover the interval between t = 1 (Aug.19, 2004), t = 573 (Nov.24. 2006).

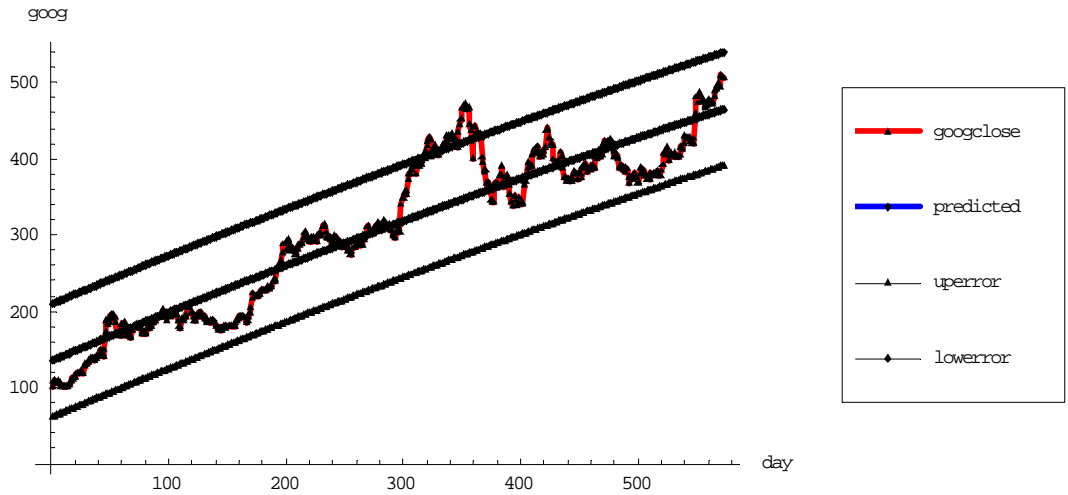

Fig.2

The cumulative distribution function, namely the probability to have a price x equal or smaller than a certain value, taking into account the starting value of 100, for linear approximation is:

$$cdf_{linear} = \frac{1}{2}\left(\mathrm{Erf}\left[\frac{0.707(x - 0.63t - 129)}{\sqrt{t}}\right] - \mathrm{Erf}\left[\frac{0.707(100 - 0.63t - 129)}{\sqrt{t}}\right]\right) \quad (15)$$

and for quadratic approximation:

$$cdf_{quadratic} = \frac{1}{2}\left(\mathrm{Erf}\left[\frac{0.707(x + 0.00013t^2 - 0.65t - 135)}{\sqrt{t}}\right] - \mathrm{Erf}\left[\frac{0.707(100 + 0.00013t^2 - 0.65t - 135)}{\sqrt{t}}\right]\right) \quad (16)$$

**Results**

The most probably day to have a price, per stock, of around 150 $, is shown in the Fig.3 and it really happened around the day no 42.

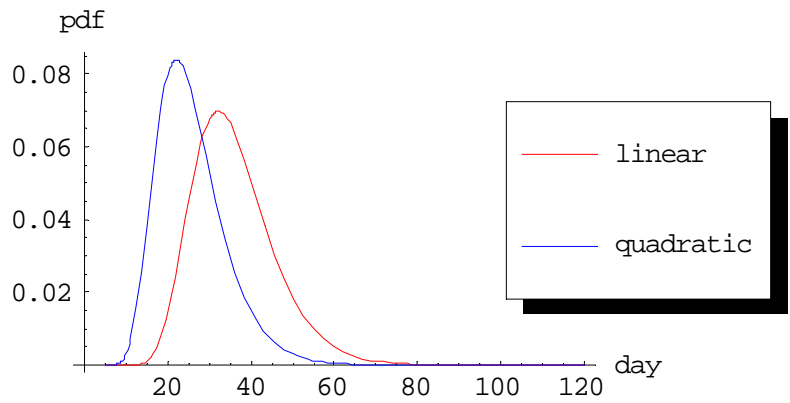

Fig.3

The corresponding cumulative distribution function is:

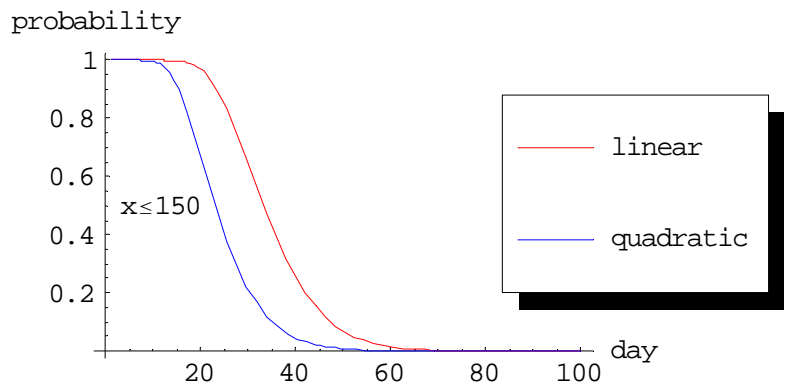

Fig.4

The most probably day to have a price, per stock, of around 250 $, is, according with this approach, shown in Fig.5, and it really happened around the day no 192.

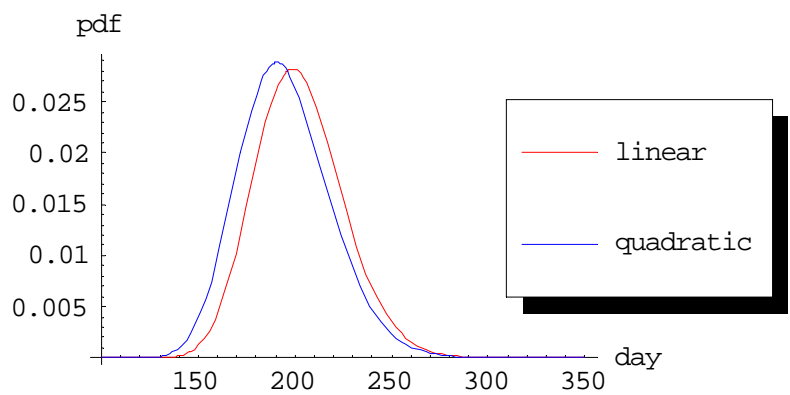

Fig.5

The corresponding cumulative distribution function is:

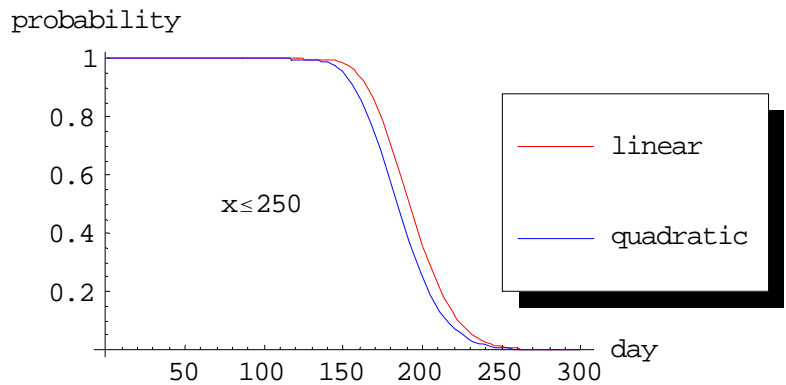

Fig.6

The most probably day to have a price, per stock, of around 500 $, is, according to this approach shown in Fig.7, and it really happened (perhaps, too early), around the day no 571.

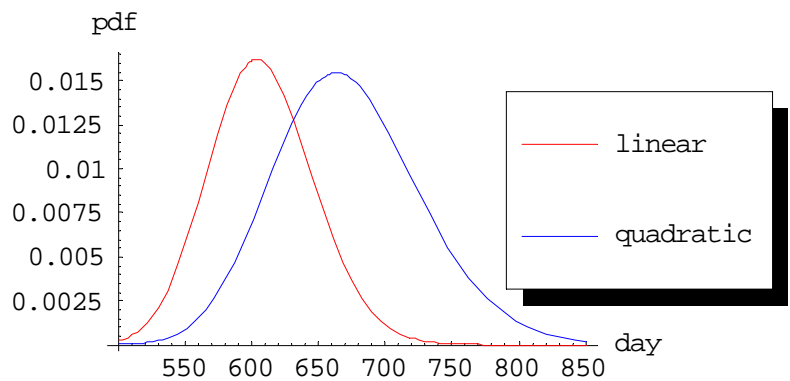

Fig.7

The cumulative distribution function corresponding to this situation is:

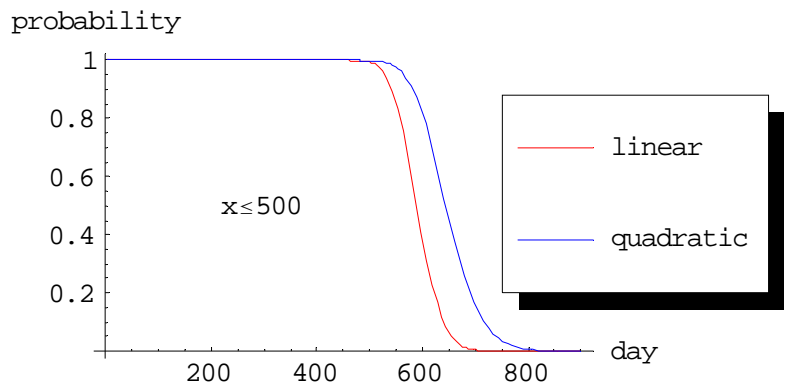

Fig.8

An estimation for a price of around 550 $ per stock gives:

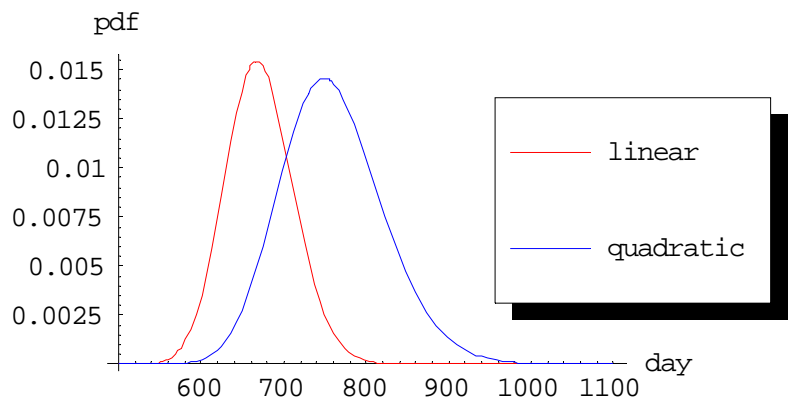

Fig.9

The corresponding cumulative distribution function being:

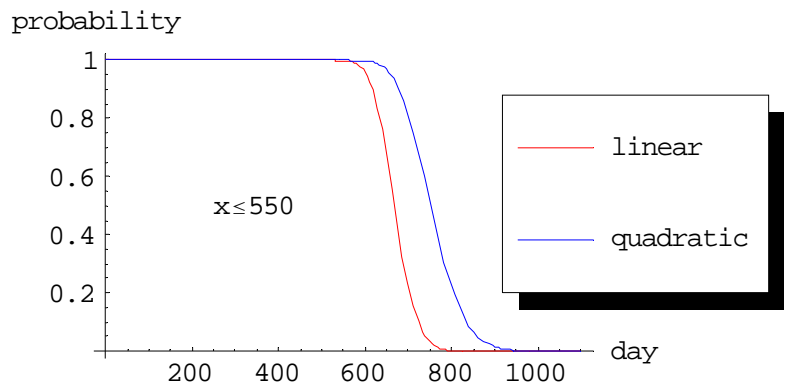

Fig.10

**Conclusions**

To apply the Fokker – Planck equation in order to study the market price evolution seems to be an appropriate one. As one can see from what was mentioned above, we cannot predict the jumps in prices, but who can? What can do such an approach is to give a reasonable idea of what probable will happen, and it for not a very long period. Data accumulation will improve the estimation.

References

[1]. H. Haken, Rev. Mod. Phys., Vol. 47, No 1, January 1977
[2]. http://finance.yahoo.com/q/hp?s=GOOG